# Precise characterization of micro rotors in optical tweezers

**Yogesha, Sarbari Bhattacharya and Sharath Ananthamurthy***

*Department of Physics, Bangalore University, Bangalore- 560 056, India.*
*\*Corresponding author: asharath@gmail.com*

We present an optical tweezer based study of rotation of microscopic objects with shape asymmetry. Thermal fluctuations and rotations are simultaneously monitored through laser back scattering. The rotation results in a modulation in intensity of the back scattered light incident on a quadrant photo detector. This results in the manifestation of peaks at a fundamental rotational frequency and at integer harmonics, superimposed on a modified Lorentzian in the power spectrum. The multiple peaks indicate that the rotations are periodic but with varying angular velocity. We demonstrate the use of video microscopy for characterization of low reflectivity rotors, such as biological cells. The methods also enable a measurement of the average torque on the rotor, and in principle, can reveal information about its principal moments of inertia, and the role of hydrodynamics at micron levels.

*OCIS codes:* 100.0100, 110.0110, 120.0120, 170.0170, 180.0180

## 1. INTRODUCTION

Microscopic objects can show rotations when optically trapped for a variety of reasons. These include rotations due to properties of the objects such as optical birefringence [1, 2] and asymmetries in shape. Rotations can also be caused due to a departure in the cylindrical symmetry of the trapping laser beam which cause unbalanced forces perpendicular to the propagation direction [3, 4]. Furthermore, angular momentum transfer from the light to the object can also result in rotations [5]. In a study of rotational dynamics of light-driven irregularly shaped gold nanorods and nanorod aggregates by Jones et al [6], a precise measurement of the speed of rotation of the trapped particles has been carried out. This measurement relies on an analysis of the correlation function of the tracked signals from the rotating particle. While such a technique enables a measurement of the laser induced fundamental rotation frequency, it can miss the possible presence of higher order harmonics if the motion is periodic but with varying angular velocity (irregular). In this work, we monitor the rotation of irregular shaped objects simultaneously with the object's thermal fluctuations in the solution. This is done through detection of the backscattered light from the object by a Quadrant Photo Detector (QPD). This yields the power spectral density of the trapped object which is a modified Lorentzian with additional peaks at specific frequencies superimposed on it. The peaks are a result of a stroboscopic modulation in the intensity of the back scattered light incident on the QPD due to the rotation. A novel feature of this recording is the revelation of additional peaks at the harmonics, apart from the one corresponding to the fundamental rotation frequency indicating that the rotation is periodic but with varying angular velocity (irregular). We extend this study to biological cells by monitoring the images obtained by Video Microscopy (VM), since the moderate or low reflectivity of light from these cells prevents formation of strong reflection maxima. The use of VM image processing in these cases is a simple and effective method to accurately measure the rotation frequency of such objects.

## 2. EXPERIMENTAL DETAILS

The optical tweezer setup used has been described elsewhere [7]. Briefly, it consists of an Ytterbium fiber laser (1064 nm, used for both trapping and tracking) guided into a 100× (1.4 NA) oil immersion objective on a customized inverted microscope. The instrument is also equipped with an XYZ stage and a QPD for detection of laser beam back scattered from the object which enables a measurement of the position of the trapped particle. In addition, there is also a provision for video imaging by a high speed CCD camera (200 frames/s, Voltrium, Singapore). For all the experiments discussed, linearly polarized light has been used and video recording of the trapped particle has been carried out simultaneously with the QPD recording of back scattered light. A few, rarely occurring, ellipsoidal shaped polystyrene beads (with approximately a long axis of length 6 µm and two short axes of length 4 µm) have been selected from a sample of spherical 3 µm polystyrene beads (Polysciences Inc., USA), dispersed in water and used in our experiments.

## 3. RESULTS AND DISCUSSION

On trapping these polystyrene beads in the optical tweezer, rotations are observed because of the laser induced torque [5, 8]. The fluctuations of the rotating bead have been tracked using the QPD through Laser Back Scattering technique (LBS). The fluctuations reveal an overall oscillatory pattern as seen in [4], with a time period same as that for the rotation of the bead. The time period of rotation is determined from an analysis of the VM recording done simultaneously. With increase in laser power, the rotational frequencies are observed to increase.

The Power Spectral Density (PSD) obtained from the QPD signal at different laser powers for the same ellipsoidal object has been shown in Fig. 1. The PSD is observed to be a modified Lorentzian, as expected for a bead in a potential well (see inset Fig. 1(ii)), but with the additional presence of peaks at integer multiples of the frequency of rotation. Such a PSD has been observed for any periodically oscillating arbitrarily shaped object with high reflectivity.

As stated earlier, the presence of peaks at higher harmonics besides the fundamental rotation frequency is because of a time variation in the angular velocity of the bead. Fig. 2 illustrates through the time dependence of the projection of the radial vector of the object, the nature of the irregular rotation.

The experimental observations are represented by a one dimensional Langevin equation modified based on the considerations that (i) the inertial term may be ignored for the low Reynolds limit [9] and (ii) apart from a thermally induced stochastic force $F_s(t)$, there is a periodic force acting on the body (a sum of a laser induced driving force as well as a response of the medium to the rotating bead) and can be written as a Fourier series.

Thus, we have [10]:

$$\gamma \frac{dx}{dt} + k\, x = F_s(t) + \sum_{n=0} a_n \cos n\Omega t \qquad (1)$$

$\gamma$ being the Stoke's drag coefficient (approximated to be the same as that of a spherical particle of diameter 3µm) and $k = 2\pi\gamma f_c$, the trap stiffness of the optical trap, $f_c$ being the corner frequency as determined from the PSD. The Fourier transform of equation (1) gives,

$$\gamma(2\pi i f)x(f) + k\, x(f) = F_s(f) + \sum_{n=0} a_n \frac{1}{2A_n}\left[\delta(f-n\Omega) + \delta(f+n\Omega)\right] \qquad (2)$$

where, $x(f) = \int_{-t/2}^{t/2} x(t')e^{2\pi i f t'} dt'$

and ($1/2A_n$) is the height of the delta function located at $n\Omega$.

Taking the complex conjugate of equation (2) and multiplying by the original, we get

$$|x(f)|^2 = \frac{\left| F_s(f) + \sum_{n=0} a_n \frac{1}{2A}\left[\delta(f-n\Omega) + \delta(f+n\Omega)\right]\right|^2}{4\pi^2\gamma^2(f_c^2 + f^2)} \qquad (3)$$

The quantity $|x(f)|^2$ is the two-sided power spectrum. The time average of $|F_s|^2$ is given by [11]

$$|F_s|^2 = 2\gamma k_B T \qquad (4)$$

where, $k_B$ is the Boltzmann constant and T, the temperature of the surrounding medium.

Hence the single-sided PSD of the bead is given by

$$S_x(f) = \frac{\left(\sqrt{2\gamma k_B T} + \sum_{n=0} B_n \delta(f - n\Omega)\right)^2}{2\pi^2 \gamma^2 (f_c^2 + f^2)} \qquad (5)$$

where, $B_n = a_n(1/2A_n)$

Equation (5) can be recast as

$$S_x(f) = \frac{\left(\sqrt{2\gamma k_B T} + \sum_{n=0} B_n \frac{1}{\sigma\sqrt{2\pi}} \exp\left(\frac{-(f-n\Omega)^2}{2\sigma^2}\right)\right)^2}{2\pi^2 \gamma^2 (f_c^2 + f^2)} \qquad (6)$$

where, the Gaussian representation of a delta function for $\sigma \to 0$ is used.

The model fits the experimentally observed data and locates the peaks at integral multiples of the rotation frequency. As the rotation frequency increases with increase in laser power the measured resonance peaks are also shifted. The observed width of peaks (0.24 Hz) is due to the finite sampling rate of the data (100 kHz).

**B. Dependence of Decay Constant on Rotation Speed**

In addition a progressive decrease in peak heights with increasing harmonic number is observed, the tallest peak being that corresponding to the fundamental frequency. The decrease in the normalized peak heights is fitted to a first order exponential decay function:

$$y = y_0 + A \exp\left(-\frac{x}{t_d}\right) \qquad (7)$$

where, $t_d$ is the decay constant. Fig. 3(a) shows the decay of peak heights as a function of the corresponding harmonics number, for different laser powers.

The decay constant plotted as a function of the laser power is shown in fig.3 (b). From the figure, it can be seen that the decay constant shows an initial increase and then a decrease with increasing power. Since the presence of harmonics is an indication of the irregularity in the rotational motion, this indicates that the motion becomes more irregular initially with increase in average angular speed but thereafter becomes less irregular. This indicates a possible role of surface hydrodynamics on the rotation, so far ignored in our analysis. Further detailed investigations are in progress to understand these effects.

### C. Measurement of Laser Induced Torque on the Rotor

In order to determine the laser induced torque on the rotating bead, the rotation drag coefficient of a prolate ellipsoid is calculated first. For a prolate ellipsoid with a long axis of length 2a and two short axes of length 2b, the rotational drag coefficient is [12]

$$\gamma_\theta = 6\eta V G_\theta \tag{8}$$

where, the volume $V=(4/3)\pi ab^2$ and $G_\theta$ is the geometric factor [12,13] characterizing the amount of deviation of the ellipsoid from a sphere. In our case, it is found to be 0.661.

For a prolate ellipsoid rotating with an angular velocity ω in a fluid with viscous coefficient η, the average drag torque is given by [14]

$$\langle \tau_{drag} \rangle = \gamma_\theta \Omega = 6\eta V G_\theta \omega \tag{9}$$

At dynamic equilibrium, average drag torque is equal to average laser induced torque and hence using equation (10), we have

$$\langle \tau_{laser\ induced} \rangle = 6\eta V G_\theta \omega \tag{10}$$

The calculated laser induced torque on the ellipsoidal bead, taking ω = 2πΩ, Ω being the fundamental frequency, has been shown as a function of laser power in fig. 4.

### D. VM Analysis

A repetition of above analysis was done from the VM recording of the bead in order to test the efficacy of a VM-based technique in analyzing the rotations. Fig. 5(a) is the two dimensional projection of the trajectory of a rotating bead formed from position tracking using an algorithm, in which the intensity center of mass of a feature (i.e. particle's local intensity maxima) is fitted assuming a Gaussian intensity distribution [15]. The corresponding one dimensional power spectral density of the rotating bead obtained is shown in fig. 5(b). Here, only peaks corresponding to the fundamental frequency of rotation and its harmonics are observed. No "knee" corresponding to the corner frequency in the PSD is seen due to the limited scan rate of the video imaging camera which restricts the monitoring to lower frequencies of the spectrum. Fig. 5(c) is an enlargement of the central marked region of fig. 5(a). This suggests a rotation pattern that can be expected from an asymmetric top. Efforts are on to model this pattern in order to extract information regarding the principal moments of inertia of the bead.

**E. Measurement of Rotational Frequency of a Trapped RBC**

As mentioned earlier, the VM based measurement technique can be effective for monitoring rotating objects with limited reflectivity from laser back scattering. We carry out these measurements on a rotating human RBC immersed in a hypertonic Phosphate Buffer Saline (PBS) solution. The higher tonicity in the medium results in the RBC shriveling from intracellular fluid loss rendering it asymmetric in shape. When trapped under linearly polarized light, the laser induced torque on the RBC results in rotations in the plane of the optical trap. Fig. 6 shows the PSD corresponding to a trapped rotating RBC. Peaks are observed at 4, 8 and 12 Hz, corresponding to the fundamental, first and second harmonics respectively.

## 4. CONCLUSIONS

We present a technique for precise measurement of rotational frequency of an optically trapped asymmetric object using detection with QPD as well as through video microscopy. Our analysis indicates that the rotations are periodic but with a time varying angular velocity, as can be surmised from the presence of peaks at the fundamental rotational frequency and additionally, at higher order integer harmonics, superimposed on a modified Lorentzian in the power spectrum. We have successfully modeled this power spectral density along with Gaussian peaks at integer multiples of the fundamental frequency of rotation. We have also been able to extend the analysis of the video microscopy to rotors with low reflectivity from laser scattering such as biological cells. In conclusion, we believe that a wealth of information with regards to the torque acting on the system, the principal moments of inertia of the rotor and the role of micro hydrodynamics in determining the rotational behavior of the system can also be extracted from such experiments.

## ACKNOWLEDGEMENT

We acknowledge a research grant from the Department of Science and Technology, Government of India (Nano Mission) that enabled this work. We thank Raghu A, Praveen P, Nagesh B V and S.A. Rangwala for useful discussions.

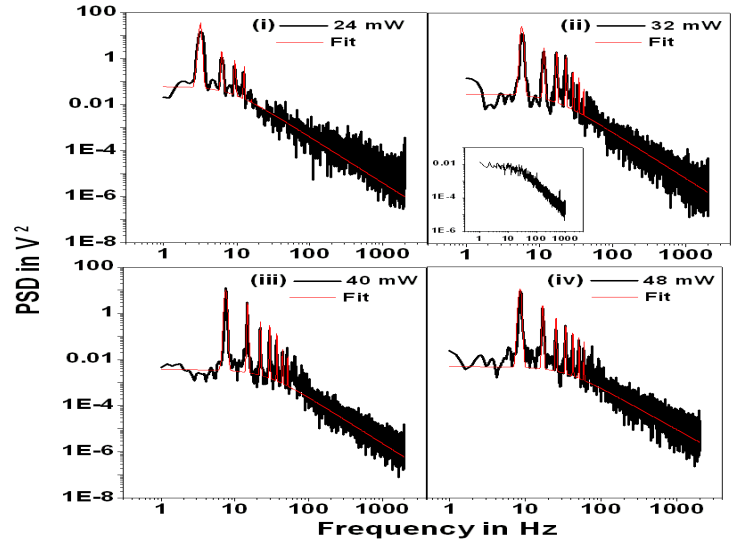

Figure 1: Power spectrum showing peaks at integral multiples of fundamental frequency for a trapped rotating bead (red lines show fits to the data).

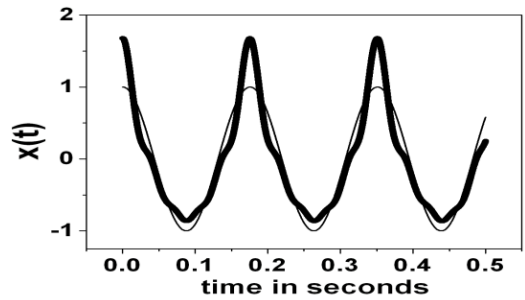

Figure 2: Variation of x(t) (projection of a radial vector of the rotating object on a fixed diametrical line) with time for a shape asymmetric rotating object at 32 mW laser power. The faint line shows the theoretically computed variation for constant angular velocity.

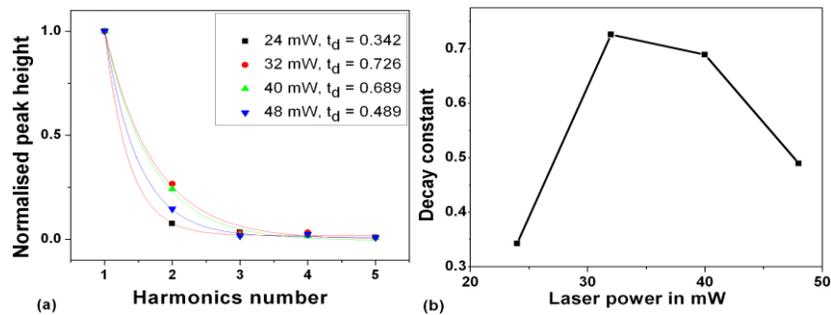

Figure 3: (a) First order exponential decay of the peak heights with the frequency at different laser powers and (b) Variation of decay constant with laser power, and hence the rotation speed.

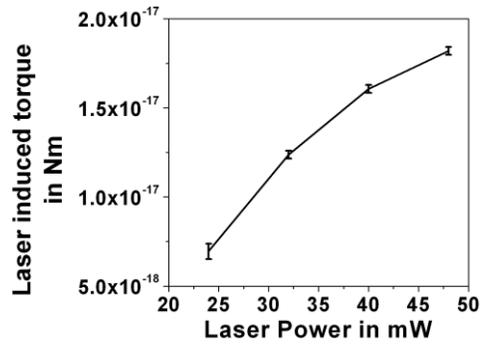

Figure 4: Variation of laser induced torque on the ellipsoidal bead as a function of laser power.

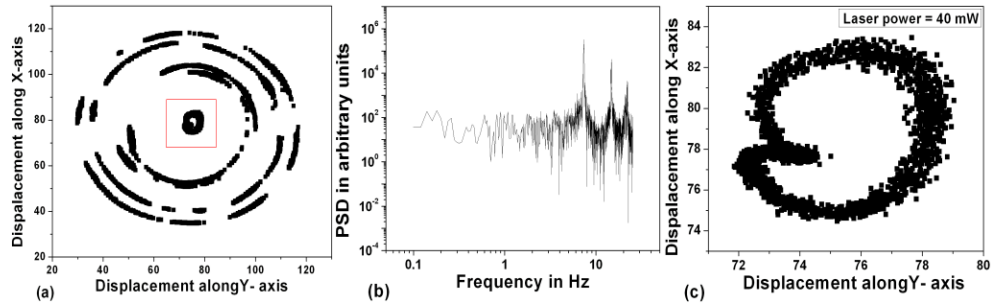

Figure 5: (a) Pattern of a rotating bead in an optical trap using image processing in VM technique. (b) Power spectral density of a rotating trapped bead showing peaks at integral multiples of frequency at a laser power of 40 mW. The peaks are observed to be at integral multiple of frequency of rotation 7.4 Hz (c) Enlargement of the central marked area of fig. 5(a).

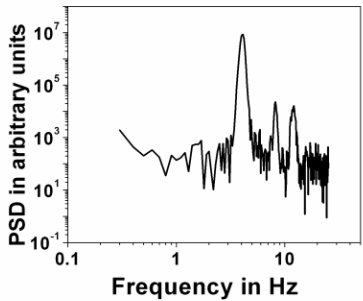

Figure 6: Power spectral density of a trapped rotating RBC showing resonance peaks at integral multiples of frequency 4 Hz, using video microscopy.